\documentstyle[prb,aps,preprint]{revtex}
\begin{document}
\draft

\title{Experimental determination of superconducting parameters for the intermetallic perovskite superconductor ${\text {MgCNi}}_3$}

%comment out next two lines for preprint, change "twocolumn" to "preprint"
%\twocolumn[
%\hsize\textwidth\columnwidth\hsize\csname@twocolumnfalse\endcsname

\author{Z. Q. Mao, M. M. Rosario, K.D. Nelson, K. Wu\cite{byline1}, I. G.
Deac\cite{byline2}, P. Schiffer, and Y. Liu\cite{byline3}}
\address{Department of Physics, The Pennsylvania State University,
University Park, PA 16802}
\author{T. He, K. A. Regan, and R. J. Cava}
\address{Department of Chemistry and Princeton Materials
Institute, Princeton University, Princeton, NJ 08540}

\date{\today}
\maketitle

\begin{abstract}

We have measured upper-critical-field $H_{\text c2}$, specific heat $C$, and tunneling spectra of the intermetallic perovskite superconductor MgCNi${}_3$ with a superconducting transition temperature $T_{\text c}\approx 7.6$~K. Based on these measurements and relevant theoretical relations, we have evaluated various superconducting parameters for this material, including the thermodynamic critical field $H_{\text c}$(0), coherence length $\xi $(0), penetration depth $\lambda $(0), lower-critical-field $H_{\text c1}$(0), and Ginsberg-Landau parameter $\kappa $(0). From the specific heat, we obtain the Debye temperature $\it \Theta _{\text D} \approx $ 280 K. We find a jump of $\Delta C/\gamma T_{\text c}$=2.3 at $T_{\text c}$ (where $\it \gamma $ is the normal state electronic specific coefficient), which is much larger than the weak coupling BCS value of 1.43. Our tunneling measurements revealed a gap feature in the tunneling spectra at $\it \Delta $ with $2\it {\Delta }/{\text k}_{\text B}T_{\text c}\approx $ 4.6, again larger than the weak-coupling value of 3.53. Both findings indicate that MgCNi$_3$ is a strong-coupling superconductor. In addition, we observed a pronounced zero-bias conductance peak (ZBCP) in the tunneling spectra. We discuss the possible physical origins of the observed ZBCP, especially in the context of the pairing symmetry of the material.

\end{abstract}

%suggested PACS numbers

%comment out bracket for preprint
%]

%bodytext

\narrowtext

\section{INTRODUCTION}
\label{sec1}

Following the remarkable discovery of superconductivity at 39~K in
MgB${}_2$\cite{1}, intermetallic compounds have received renewed interest in the search for new superconducting materials. A new intermetallic superconductor, MgCNi${}_3$, with the superconducting transition temperature $T_{\text c}\approx 7.6$~K was found recently by He {\it et al} \cite{2}.  This compound has a cubic perovskite crystal structure, with Ni, C, and Mg replacing O,
Ti and Sr, for example, in the more familiar oxide perovskite SrTiO$_3$. It is a unique material that bridges the intermetallic compound MgB$_2$ and perovskite oxide superconductors.

The electronic structure of MgCNi${}_3$, calculated by several groups with different methods \cite{3,4,5,6,7}, indicated that the electronic states of MgCNi${}_3$ at the Fermi energy ($E{}_F$) are dominated by the 3$d$ orbitals of Ni. The derived density of states (DOS) has a sharp peak just below $E{}_F$, which leads to a moderate Stoner enhancement. This signals the presence of substantial ferromagnetic (FM) spin fluctuations in MgCNi$_3$ \cite{6,7}, as confirmed by $^{13}$C NMR measurements \cite{8}. The presence of such FM fluctuations may favor unconventional pairing as in Sr$_2$RuO$_4$ \cite{7,9}. However, the nuclear spin-lattice relaxation rate 1/$T_{\text 1}$ obtained in NMR measurements \cite{8} exhibited a clear coherence peak just below $T_{\text c}$, typical for an isotropic $s$-wave superconductor. 

We measured the temperature dependence of resistivity in various magnetic field, and obtained the upper-critical field $H_{\text c2}(T)$ for this material. We also carried out the specific-heat measurements over a wider temperature region of 20-0.3~K on a high-quality MgCNi$_3$ sample. The value of $T_{\text c}$ for this sample is 7.63 K as opposed to 6.2 K for the sample used in the previous measurements \cite{2}. From these measurements, we estimated various superconducting parameters of MgCNi$_3$, including the thermodynamic critical field $H_{\text c}$(0), coherence length $\xi $(0), penetration depth $\lambda $(0), lower-critical-field $H_{\text c1}(0)$, and Ginzberg-Landau (GL) parameter  $\kappa $(0). 

The superconducting energy gap ($\it \Delta $) of MgCNi$_3$ has been estimated from NMR measurements \cite{8}, yielding a value of $2\it \Delta /{\text k}_{\text B}T_{\text c}\approx $ 3.2 where $T_{\text c}$ is the superconducting transition temperature. This value is less than the weak-coupling value of 3.53. However, our specific heat measurements revealed a jump of $\Delta C/ \gamma T_{\text c}$ = 2.3 (1.9 in Ref. 2) at the $T_{\text c}$ of 7.63 K, considerably larger than the weak-coupling value 1.43, which is inconsistent with the NMR result. 

To determine the superconducting gap directly, as well as to obtain information on the pairing symmetry of this new superconductor with apparently large ferromagnetic fluctuation similar to Sr$_2$RuO$_4$, we have carried out tunneling spectroscopy measurements on high-quality polycrystalline MgCNi$_3$ samples. From these measurements, we found a gap feature with the ratio of $2{\it \Delta }/{\text k}_{\text B}T_{\text c}\approx 4.6$. This result, together with the specific heat data, suggests that MgCNi$_3$ is a strong-coupling superconductor. In addition, we observed a pronounced zero-bias conductance peak (ZBCP) in tunneling spectra. We discuss its possible physical origins, especially in the context of the pairing symmetry of the material.

\section{EXPERIMENTAL}
\label{sec2}

The polycrystalline material used for this study was synthesized by
solid-state reaction. The preparation details can be found in Ref. [2].  
We determined $H_{\text c2}(T)$ from four-probe resistance measurements. Specific heat measurements were performed using a relaxation method in a commercial calorimeter in a Quantum Design Physical Property Measurement System.  

We prepared mechanical junctions for tunneling spectrum measurements. They were made on polycrystalline samples using a sharp W tip with an end radius of about 15~$\mu $m. The sample and the W tip were held together by an insulated Cu frame, with the tip touching the sample gently. The sample surface was examined by scanning electron microscopy. The average grain size on the surface was found to be about 5~$\mu $m. As a result, the mechanical tunnel junction involved only a few MgCNi$_3$ grains.

Current-voltage ($I-V$) characteristics, as well as junction resistance,
was measured by a four-point method in a 1 K pot dip probe and a $^3$He cryostat. To reduce heating, the $I-V$ curves were measured by a dc pulsed-current method with a typical pulse duration of 50 ms followed by a 2 s delay between successive pulses. The tunneling conductance ($G = {\text d}I/{\text
d}V$) was computed numerically from the measured $I-V$ curves.

\section{RESULTS AND DISCUSSIONS}
\label{sec3}

\subsection{Electrical transport}
\label{subsec1}
 
Figure \ref{1} shows the $H-T$ phase diagram obtained from the $R$ vs. $T$
curves at different fields. Here $T_{\text c}$ is defined as the
intersection of the linear extrapolation of the most rapidly changing part
of $R(T)$ and that of the normal state resistance, as shown in the inset
of Fig. \ref{1}. Within the weak-coupling BCS theory, $H_{\text c2} (T=0)$ can
be estimated using the Werthamer-Helfand-Hohenberg (WHH) formula\cite{10},
	\begin{equation}
	$$\mu _0H_{\text c2}(0) = -0.693({\text d}H_{\text c2}/
	{\text d}T)_{T=T_{\text c}}T_{\text c}$$
	\label{eq1}
	\end{equation}
which leads to a $\mu _0 H_{\text c2}(0)$ value of 14.8 T.  Meanwhile the
Pauli-limiting field
	\begin{equation}
	$$\mu _0H^{\text {Pauli}} = 1.24{\text k}_{\text B}T_{\text c}/
	\mu _{\text B}$$
	\label{eq2}
	\end{equation}
expected within the same weak-coupling BCS theory\cite{11} is about 14 T
for $T_{\text c}=7.60$~K. The $\mu _0H^{\text {Pauli}}$ is about 0.8 T less than the $\mu _0H_{\text c2}(0)^{\text {WHH}}$, suggesting that pair-breaking effects due to the Zeeman energy in MgCNi${}_3$ is small. From our experimental results of specific heat and tunneling spectra, to be presented below, MgCNi${}_3$ appears to be a strong-coupling superconductor. Taking into account the effects of strong coupling, a reasonable extrapolation of $H_{\text c2}(0)$ should lie between $\mu _0H^{\text {Pauli}}$ and $H_{\text c2}(0)^{\text {WHH}}$, around 14.4 T, based on which the superconducting coherence length $\xi (0)$ can be estimated to be approximately 46~\AA, using the GL formula for an isotropic three-dimensional superconductor, $\mu _0H_{\text c2}(0)= \phi _0/2\pi \xi (0)^2$. This value is in good agreement with a earlier report of $\xi (0)= 47$~\AA. \cite{12}

\subsection{Specific heat}
\label{subsec2}

We measured the specific heat of MgCNi$_3$ from 20 K down to 0.4~K. As shown in the inset of Fig. \ref{2}, the specific-heat data in the temperature region of $T_{\text c}^{\text onset}(\approx 9.4$~K) $< T < 16$~K, was fitted well with the form $C(T)={\gamma }T + {\beta }T^3$, which is the usual temperature dependence of specific heat for metals at $T <<\it \Theta_{\text D}$. The $\it \Theta_{\text D}$ can be determined from the coefficient of the $T^{\text 3}$ term $\beta = N(12/5)\pi^{\text 4}R\it \Theta_D^{-3}$, where $R$=8.314 J/mol-K, and $N$=5 for MgCNi$_3$. The linear term of $C(T)$ is due to the electronic contribution, while the $T^{\text 3}$ term comes from the phonon contribution. From the above fitting, we obtained $\gamma $=9.2 mJ/mol-Ni.K$^2$ and $\it \Theta_D \approx $ 280~K.

By subtracting the phonon contribution from the $C(T)$, we obtained the temperature dependence of electronic specific heat $C_{\text e}(T)$. The main panel of Fig. \ref{2} shows $C_{\text e}/T$ as a function of $T$. It reveals a much sharper superconducting transition than seen in the specific-heat data in Ref. 2. From an entropy-conserving construction (see the solid lines in the figure), the midpoint transition temperature $T_c^{\text mid}$ was estimated to be 7.63~K, and the specific-heat jump $\Delta C/T_c^{\text mid} =  21.1$~mJ/mol-Ni K$^2$. This value is considerably larger than that reported in Ref. 2 where $\Delta C/T_c^{\text mid}=19.0$~mJ/mol-Ni.K$^2$ with $T_{\text c}^{\text mid}=6.2$~K, indicating that the sample used in the present study is of improved quality. On the other hand, we note that the $C_{\text e}/T$ shows an unusual behavior at low temperature, i.e., an upturn for $T< 2$~K. This upturn is most likely due to the paramagnetism of unreacted Ni impurities. Similar behavior was observed in MgB$_2$ where Fe impurities led an upturn in $C/T$ at low temperature \cite{13}.

It is known that the parameter $\Delta C(T_{\text c})/\gamma T_{\text c}$ can be used to measure the strength of the electron coupling \cite{14}. In the BCS weak-coupling limit, its value is 1.43. For MgCNi$_3$, the ratio of $\Delta C(T_{\text c})/\gamma T_{\text c}$ estimated from the above data is approximately 2.3 (1.9 in Ref. 2), much larger than the weak coupling value. This indicates that MgCNi$_3$ is a strong-coupling superconductor with coupling strength close to that of Hg with $\Delta C(T_{\text c})/\gamma T_{\text c} = 2.37.$ This is consistent with our results of tunneling measurements (see below).

Based on the specific data and $H_{\text c2}(0)$ obtained above, we can estimate various superconducting parameters using relevant theoretical relations for isotropic 3D superconductor. The $H_{\text c}$(0) can directly be estimated by integrating the specific data of Fig. \ref{2} in the superconducting state and using the relation:
	\begin{equation}
	\mu_{\text 0}H_{\text c}(0)^{\text 2}/2 = -\gamma T_{\text c}^{\text 2}/2      	+ \int _0^{T_{\text c}}C_{\text e}(T)dT.
	\label{eq3}
	\end{equation}
The $\mu _{\text 0}H_{\text c} (0)$ thus evaluated is 0.22~T. The GL parameter $\kappa $ was evaluated from
	\begin{equation}
	H_{\text c2}(0)= \sqrt 2\kappa H_{\text c}(0)
	\label{eq4}
	\end{equation}
to be 46, which is quite large, but not as large as that of high-$T_{\text c}$ superconductors \cite{15}. From $\kappa = \lambda /\xi $, the estimated penetration depth is $\lambda (0) = 2130 $~\AA . Using 
	\begin{equation}
	H_{\text c1}(0)H_{\text c2}(0) = H_{\text c}(0)^{\text 2}(\text 	{ln}\kappa (0) + 0.08)
	\label{eq5}
	\end{equation}
valid for $\kappa >>$ 1, the lower critical field $H_{\text c1}$ was estimated to be 13 ~mT. All estimated parameters are summarized in Table 1. 

\subsection{Single-particle tunneling}
\label{subsec3}

Figure ~\ref{3} displays a set of typical tunneling conductance spectra ($G$ vs. $V$) of MgCNi$_3$ (Junction MCN/W \#6). The junction resistance, $R_{\text J}(T)$, measured at 0.4~mA (corresponding to a voltage below 0.02~mV for $T<8$~K), started to drop at the bulk $T_{\text c}$, 7.63~K, as shown in the left inset where bulk resistivity is also included for comparison. This drop was much broader than the transition seen in bulk resistivity, suggesting that the junction resistance must have been dominated by the tunnel barrier. Therefore the voltage drop should primarily occur at the junction interface.

As seen in the main panel of Fig. \ref{3}, the dominant feature, a pronounced zero-bias conductance peak (ZBCP) accompanied by two dips, starts to develop around the bulk $T_{\text c}$. The dip feature had a smooth, clearly identifiable temperature dependence up to bulk $T_{\text c}$. Therefore, the energy where $G$ shows a sharp drop (as indicated by arrows in Fig. \ref{3}), is clearly a characteristic energy $E_{\text c}$ associated with the bulk phase of MgCNi$_3$. If we identify this energy as the superconducting gap ${\it \Delta }$, we estimated the $\it \Delta $(0) for MgCNi$_3$ to be 1.5~meV, with  2${\it \Delta }/k_{\text B}T_{\text c}\approx $ 4.6. This ratio is larger than the BCS weak-coupling value, 3.53, but close to that of strong coupling superconductor Hg (2${\it \Delta }/k_{\text B}T_{\text c}\approx  4.3$), consistent with the indication obtained from the specific-heat measurement. The right inset of Fig. \ref{3} shows the temperature dependence of $E_{\text c}$ for this junction. For comparison, we include the ${\it \Delta (T)}$ in the weak-coupling BCS model into this inset (dashed line). Clearly the $E_{\text c}(T)$ obtained as described above is higher than the ${\it \Delta (T)}$ of weak-coupling BCS at lower temperatures, again supporting the statement that MgCNi$_3$ is a strong-coupling superconductor.

Figure \ref{4} shows the tunneling spectra of another junction (MCN/W \#4) measured down to 0.4~K. The tunneling features seen in this junction look somewhat different from junction MCN/W \#6 shown in Fig.\ref{3}: the conductance peak was much broader and did not split at higher temperatures ($>3$~K); but evolved into a spectrum of sharper central ZBCP with two side peaks as $T< 3$~K. We believe that such a difference is most likely caused by distinct barrier strength at the junction interface. Nevertheless, the characteristic energy $E_{\text c}$ defined using the same way as described above (see the arrows in Fig.\ref{4}) seems comparable to that defined in junction MCN/W \#6.

It is interesting to consider the possible physical origins of the ZBCPs observed in the tunneling spectra. First, it is known that magnetic impurities in or near the barrier can lead to a ZBCP due to magnetic and Kondo scattering (the Appelbaum-Anderson model) \cite{16}. For the MgCNi${}_3$ samples used in the present study, a small amount of unreacted Ni was very likely to exist, as reflected in the specific-heat measurement. It might appear near the tunnel barrier as a magnetic impurity. However, a ZBCP due to magnetic impurities should be uncorrelated with the occurrence of superconductivity. In contrast, the ZBCP in the present case was found to emerge just below bulk $T_{\text c}$. In addition, the height of the ZBCP caused by magnetic impurities should depend on temperature logarithmically \cite{16}, which was not found in our experiment. Further, in the Appelbaum-Anderson model, a magnetic field should split the ZBCP. But in our experiment, we found that the ZBCP became sharper under a magnetic field, instead of splitting, as shown in the right inset of Fig. \ref{4}. Therefore, magnetic impurities cannot be responsible for the observed ZBCP in MgCNi$_3$.

Although our tunnelling spectra were obtained at a superconductor/normal-metal  (S/N) interface, the observed ZBCP cannot be due to conventional Andreev reflection. The values of $G_{\text s}/G_{\text n}(V=0)$ ($G_{\text n}$ is the normal state zero-bias conductance at 8 K), which were found to be around 5 in these two junctions discussed above, were considerably larger than the maximum value of 2 expected for the conventional Andreev reflection \cite{17}. 

Another mechanism for the ZBCP that is particularly relevant to our experiment is related to Josephson-coupling effects. The junctions we used for this study were prepared on polycrystalline samples. These samples could consist multiple junctions formed between W tip and MgCNi$_3$ grains. Each of these grains was in turn connected with other MgCNi$_3$ grains, forming parallel conduction channels. Each channel would then consist a W/MCN junction and additional MCN/MCN junctions in series. The Josephson coupling at a dominating intergrain interface may result in a large ZBCP in the tunneling spectrum, as observed in tunnel junctions prepared on polycrystalline MgB$_2$ \cite{18}. However, the ZBCP originating from intergrain interface should grow rapidly in height with decreasing temperature because of the emergence of Josephson coupling at low temperatures. Our observation of the saturation in the height of d$I$/d$V$($V=0$) below 2.0~K (see the left inset of Fig.\ref{4}) is not consistent with such a simple intergrain-coupling picture. To completely rule out the intergrain coupling as the origin of the observed ZBCP, more experiments on single crystal samples, which have not been available yet, are certainly necessary.

Can the observed ZBCP be due to the surface mid-gap Andreev Bound states (ABSs) \cite{19,20} of a non-$s$-wave superconductor? ABSs arise from quantum interference during the Andreev reflection at the surface of an unconventional superconductor where the scattered quasi-particles experience a phase change. ABSs manifest as a ZBCP, as observed in $d$- and $p$-wave superconductors \cite{21,22,23,24}. To examine if there is such a possibility for our ZBCP observations, we have estimated the temperature dependence of spectral weight of central ZBCP and side peak in Fig. \ref{4} by integrating $G/G_{\text n}$ in the appropriate ranges of $V$. It was found that a clear spectral-weight transfer from the side peak to the ZBCP occurs with decreasing temperature. This suggests that the ABSs is a possible origin for the ZBCP observed in MgCNi$_3$. Therefore we should not exclude the possibility of unconventional pairing in MgCNi${}_3$. Our suggestion seems consistent with a recent result of band calculations that indicates that MgCNi$_3$ is near a ferromagnetic instability and its superconductivity is unconventional in nature \cite{7}.

However, as pointed earlier, NMR measurements revealed a coherence peak in 1/$T_{\text 1}$ just below $T_{\text c}$ \cite{8}, which appears to suggest that MgCNi$_3$ is an s-wave superconductor. This seems to contradict to our observation and the theoretical indication \cite{7}. More experiments are clearly needed to reconcile the apparently opposite indications offered by these two types of measurements. If the pairing symmetry in MgCNi$_3$ turns out to be indeed $s$-wave, then our observation of the anomalous behavior in the ZBCP will still be of interest as it demands a new model for the ZBCP within the framework of $s$-wave superconductivity. 

Finally, we comment on why the coupling strength estimated from NMR \cite{8} is inconsistent with that reflected in specific-heat and tunneling measurements.  The ratio of 2${\it \Delta }/k_{\text B}T_{\text c}$(=3.2) evaluated by NMR was obtained from the exponential fit of the temperature dependence of 1/$T_{\text 1}$ (measured at 0.45 T). In principle, the most reliable $\it \Delta $ value should be obtained by the fit of 1/$T_{\text 1}$ to temperatures much lower than $T_{\text c}$ to avoid the crossover regime. However, the 1/$T_{\text 1}$ fit in Ref. [8] was extended only down to 2.5 K.  Below this temperature, the 1/$T_{\text 1}$ became saturated. This saturation behavior was ascribed to the flux avalanche effect. If the diminished temperature range resulted in a low estimate of $\it \Delta $, then these NMR data and the present specific-heat and tunneling results might be reconciled.

\section{CONCLUSION}
\label{sec4}

In conclusion, we have carried out upper-critical field, specific heat and tunneling spectroscopy measurements of the newly discovered intermetallic perovskite superconductor MgCNi${}_3$. We have evaluated various superconducting parameters using the specific heat data and $H_{\text c2}$(0). The specific heat measurement gives $\Delta C/\gamma T_{\text c} = 2.3$ and the Debye temperature, $\it \Theta_D \approx $ 280~K, indicating that MgCNi$_3$ is a strong-coupling superconductor. We have also seen a corresponding signature of strong coupling in the tunneling measurement of MgCNi$_3$, which reveals a gap feature with the ratio of $2\it \Delta /{\text k}_{\text B}T_{\text c}\approx $ 4.6. In addition, from the presence of a ZBCP in tunneling spectra of MgCNi$_3$, we suggest that the possibility of unconventional pairing in MgCNi$_3$ should not be excluded even though the NMR experiment suggests that it is a $s$-wave superconductor. Additional experimental probes, as well as single crystal samples, are needed to further elucidate the pairing symmetry of this novel superconductor.

We would like to thank T. Imai, D.J. Singh, I.I. Mazin and N-C. Yeh for useful discussions. This work was supported at Penn State by NSF through Grants No. DMR-9974327, DMR-0101318 and DMR-9702661, and at Princeton by NSF grants DMR-9809483 and DMR-9725979, and DOE through grant DE-FG02-98-ER45706.

%references 

%figures

\begin{figure}
\caption{Upper critical field $\mu _0H_{\text c2}$ as a function of
temperature for MgCNi${}_3$. The square symbol is $\mu _0H_{\rm c2}(0)$
estimated using WHH theory and the triangle is the estimated
Pauli-limiting field (see text).  The inset shows the temperature
dependence of sample resistance in zero field.}
\label{1}
\end{figure}

\begin{figure}
\caption{Temperature dependence of the electronic specific heat divided by temperature, $C_{\text e}/T$, of MgCNi$_3$. The inset shows the $C/T$ as a function of $T^2$ above $T_{\text c}$. The solid line is the fit of the experimental data to $C = \gamma T + \beta T^{3}$ between 9.5 and 16~K.}   
\label{2}
\end{figure}

\begin{figure} 
\caption{ Tunneling spectra of $G$=d$I$/d$V\sim V$ (on a logarithmic scale) for junction MCN/W \#6 at different temperatures. All curves except the top curve have been shifted downwards for clarity by multiplying $G$ with a numerical factor. The left inset shows the junction resistance $R_{\text J}(T)$ and bulk resistivity. The right inset shows the temperature dependence of the characteristic energy  $E_{\text c}$, as well as the BCS ${\it \Delta (T)}$.} 
\label{3} 
\end{figure}

\begin{figure}
\caption{ Tunneling conductance spectra (on a logarithmic scale) for junction MCN/W \#4. All curves except the top curve have been shifted downwards for clarity by multiplying $G$ with a numerical factor. The left inset shows the $R_{\text J}(T)$ of junction MCN/W \#4, measured at 1mA (corresponding to a voltage below 0.08 meV for $T<$ 8 K). The right inset shows tunneling conductance spectra of junction MCN/W \#4 measured under fields of 0~T and 0.5~T at 1.5~K.} 
\label{4}
\end{figure}

\begin{table}
\caption{Characteristic parameters of intermetallic perovskite superconductor MgCNi$_3$}

\label{table1}

\begin{tabular}{c}

Parameters\\ 
\tableline
$T_{\text c} = 7.63$~K\\
$\mu _{\text 0}H_{\text c2} (0) = 14.4$~T\\
$\mu _{\text 0}H_{\text c} (0) = 0.22$~T\\
$\mu _{\text 0}H_{\text c1} (0) = 13$~mT\\
$\xi (0) = 46$~\AA\\
$\lambda (0) = 2130$~\AA\\
$\kappa (0) = 46$\\ 
$\gamma = 9.2$~mJ/mol-Ni.K$^2$\\
$\it \Theta _{\text D} = $280~K\\
$\Delta C/\gamma T_{\text c} = 2.3$\\
$2\it \Delta /{\text k}_{\text B}T_{\text c}=$ 4.6\\
\end{tabular}

\end{table}

\end{document}